\documentclass[apj]{emulateapj}
\usepackage{apjfonts}
\usepackage{amsbsy}

\def\hmpc{$h^{-1}$Mpc}
\def\hkpc{$h^{-1}$kpc}

\def\hmsol{$h^{-1}$M$_\odot$}
\def\kms{km\,s$^{-1}$}

\def\kmsmpc{km\,s$^{-1}$\,Mpc$^{-1}$}

\def\om{\Omega_m}

\def\s8{\sigma_8}

\def\rhocrit{\rho_{\rm crit}}

\def\x2{$\chi^2$}
\def\hmsol{$h^{-1}\,$M$_\odot$}

\def\NNm1{\langle N(N-1) \rangle}

\def\m_star{M_\ast}

\def\rhob{\tilde{\rho}}

\def\om{\Omega_m}

\def\s8{\sigma_8}

\def\hmpc{$h^{-1}\,$Mpc}
\def\hkpc{$h^{-1}\,$kpc}

\def\x2{$\chi^2$}
\def\hmsol{$h^{-1}\,$M$_\odot$}

\def\kms{km\,s$^{-1}$}
\def\kmsmpc{km\,s$^{-1}$\,Mpc$^{-1}$}

\def\NNm1{\langle N(N-1) \rangle}

\def\p0{P_0(r)}

\def\rhob{\rho_b}
\def\cg{\kappa_g}
\def\pb{\rho_b}

\def\mtwelve{M_{12}}

\def\rshock{R_{\rm sh}}
\def\rg{R_g}%{R_{\rm gas}}
\def\rgas{R_g}%{R_{\rm gas}}
\def\rgstar{R_{g\ast}}

\def\ashock{\gamma_{\rm sh}}%{\alpha_{\rm shock}}
\def\logmbar{\langle \log M\rangle}
\def\fcold{f_{\rm cold}}
\def\wr{W_r}
\def\fr{f(W_r)}
\def\rs0{\hat{R}_{\rm sh}^0}
\def\mg2{Mg\,II}

\def\mcrit{M_{\rm crit}}

\def\rhobar{\bar{\rho}_m}
\def\lstar{L_\ast}

\def\mgdot{\dot{M}_g}

\def\ktot{\kappa_{\rm tot}}

\begin{document}

\title{On the Redshift Evolution of Mg\,II Absoprtion Systems}

\author{ Jeremy L. Tinker$^1$ \& Hsiao-Wen Chen$^2$ }
\affil{$^1$Berkeley Center for Cosmological Physics, University of California-Berkeley\\
  $^2$ Department of Astronomy and Astrophysics \& Kavli Institute for
  Cosmological Physics, University of Chicago}

\begin{abstract}

  We use a halo occupation approach to connect \mg2\ absorbers to dark
  matter halos as a function of redshift. Using the model constructed
  in \cite{tinker_chen:08}, we parameterize the conditional
  probability of an absorber of equivalent width $\wr$ being produced
  by a halo of mass $M_h$ at a given redshift, $P(\wr|M_h,z)$. We
  constrain the free parameters of the model by matching the observed
  statistics of \mg2\ absorbers: the frequency function $\fr$, the
  redshift evolution $n(z)$, and the clustering bias $b_W$. The
  redshift evolution of $\wr\ge 1$ absorbers increases from $z=0.4$ to
  $z=2$, while the total halo cross section decreases monotonically
  with redshift. This discrepancy can only be explained if the gaseous
  halos evolve with respect to their host halos. We make predictions
  for the clustering bias of absorbers as a function of redshift under
  different evolutionary scenarios, eg, the gas cross section per halo
  evolves or the halo mass scale of absorbers changes. We demonstrate
  that the relative contribution of these scenarios may be constrained
  by measurements of absorber clustering at $z\gtrsim 1$ and $z\sim
  0.1$. If we further assume a redshift-independent mass scale for
  efficient shock heating of halo gas of $\mcrit = 10^{11.5}$ \hmsol,
  absorber evolution is predominantly caused by a changing halo mass
  scale of absorbers. Our model predicts that strong absorbers always
  arise in $\sim \mcrit$ halos, independent of redshift, but the mass
  scale of weak absorbers decreases by 2 dex from $0<z<2$. Thus, the
  measured anti-correlation of clustering bias and $\wr$ should
  flatten by $z\sim 1.5$.

\end{abstract}

\keywords{cosmology: observations---intergalactic medium---quasars:
  absorption lines---dark matter: halos}

\section{Introduction}

Absorption lines in the spectra of quasars reflect the presence of gas
intersected along the lines of sight to the QSOs. A subset of these
absorption lines systems, i.e., \mg2, C\,IV, Lyman-limit systems, are
created when a line of sight passes through a dark matter halo that
contains cool or warm gas. The presence of Mg$^+$ indicates that the
gas is photoinized with a temperature near $10^4$ K. The connection
between \mg2\ absorbers and dark matter halos is well established:
\mg2\ absorbers correspond to neutral hydrogen column densities of
$10^{18}-10^{20}$ cm$^{-2}$ (\citealt{churchill_etal:00,
  rao_etal:06}), densities that exist in collapsed objects. Blind
surveys of galaxies near QSO sightlines yield \mg2\ absorbers a high
fraction of the time (\citealt{tripp_bowen:05, chen_tinker:08}), and
known \mg2\ absorbers are nearly always found to be associated with a
galaxy with a projected separation significantly less than the virial
radius of the dark matter halo around that galaxy (e.g.,
\citealt{bergeron:86, steidel:93, kacprzak_etal:08}). \mg2\ absorbers
thus probe the cool gas around galaxies from which these galaxies form
their stars. The redshift evolution of absorbers is thus connected
both to the growth of dark matter structures and the evolution of gas
accretion in dark matter halos. Connecting quasar absorbers with dark
matter halos is a direct probe of the gas accretion history of such
objects.

The observed statistics of \mg2\ absorbers are a consequence of the
properties of the halos in which the gas resides.  In this paper we
present a model that provides a mapping between rest-frame \mg2\
absorption equivalent width, $W_r(2796)$ (hereafter $\wr$) and dark
matter halo mass $M_h$ in such a way as to match the frequency
function of absorbers, their large-scale clustering bias, and the
redshift evolution of their number counts.

The methods for quantifying the connection of galaxies to dark matter
halos are now well established through halo occupation models (see,
e.g., \citealt{seljak:00, roman_etal:01, cooray_sheth:02,
  berlind_weinberg:02} for early works and \citealt{zheng_etal:07,
  vdb_etal:07, tinker_etal:07_pvd} for examples of more recent
extensions of the analysis). In \cite{tinker_chen:08} (hereafter Paper
I), we presented a model of the halo occupation of cold gas that
addressed the first two of these issues: the frequency and bias
functions at $z=0.6$. Our approach in Paper I was to parameterize the
conditional probability of an absorber being produced by a halo of
given mass $P(\wr|M_h)$ and using the statistics of dark matter halos
to constrain the free parameters of $P(\wr|M_h)$ in order to match the
observations. After adopting a cored isothermal density profile for
the \mg2\ gas (a choice later supported in \citealt{chen_tinker:08},
hereafter Paper II), the two free quantities governing $P(\wr|M_h)$
were constrained from the data: the incidence of absorption per halo
$\cg(M_h)$ and the absorption efficiency $A_W(M_h)$. The former
encapsulates both the covering fraction of \mg2\ gas in the dark
matter halo and the incidence of extended gaseous halos. The latter
controls the mean absorption strength as a function of halo
mass. Because the halo-based model parameterizes these two quantities,
it makes no assumptions about the physical nature of these
systems---whether they are produced from infalling or outflowing gas,
or within a turbulent, static gas halo. The primary purpose of this
model is to determine the distribution of halo masses probed by \mg2\
absorbers and the incidence required to match observations.

The difficulty in reconciling halo statistics with measured statistics
of \mg2\ absorbers at $z=0.6$ lies in the observed anti-correlation
between $\wr$ and the large-scale clustering bias $b_W$
(\citealt{bouche_etal:06, gauthier_etal:09, lundgren_etal:09}). For
dark matter halos, the most biased objects are the rarest and most
massive. The anti-correlation between $\wr$ and $b_W$ implies that the
most biased absorption systems are the weakest, most frequent
systems. In Paper I we concluded that the $\fr$ and $b_W$ data could
only be fit by a model that contained a critical mass scale below
which absorption efficiency is high, but at higher masses the
absorption efficiency is significantly attenuated due to a lack of
cold gas in the halo. This transition from `cold-mode' gas accretion
to `hot-mode' accretion is motivated by results of analytic models and
numerical simulations that demonstrate a critical halo mass at which
shock heating becomes efficient (\citealt{birnboim_dekel:03,
  keres_etal:05, keres_etal:08, dekel_birnboim:06, birnboim_etal:07,
  brooks_etal:08}). These simulations found that the transition
between the cold mode and hot mode occurs between $10^{11}$ \hmsol\
and $10^{12}$ \hmsol, precisely the location required to match the
\mg2\ statistics. The expectations of these papers are qualitatively
similar to those from two-phase gas cooling models of
\cite{mo_jordi:96} and \cite{maller_bullock:04}. In these models, in
halos below $\sim 10^{11.5}$ \hmsol\ gas cooling is rapid, while at
higher masses the fraction of cold gas in the halo (in the form of
pressure-supported cold clouds) decreases exponentially with halo
circular velocity, leaving only a small fraction of cold gas at
$M_h\gtrsim 10^{13}$ \hmsol. These observational trends may also be
produced if the distribution of cold gas changes within the halo as
well; the cold gas fraction may remain constant but be redistributed
into a smaller cross section within dense gas clumps or be less
extended in the dark matter halo. Along with a high-mass cutoff, the
data also require a low-mass cutoff below $M\sim 10^{11}$ \hmsol\ due
to the high overall bias of absorbers. The theoretical motivation for
such a cutoff is less clear than a high-mass cutoff, and we will
discuss both equally in this paper.

Paper I demonstrated that the anti-correlation between $\wr$ and $b_W$
is produced from residual amounts of cold gas in hot-mode halos. This
small amount of cold gas can produce only low equivalent width
absorption in $M_h\gtrsim 10^{13}$ \hmsol\ halos, increasing the bias
of low-$\wr$ systems. This scenario is similar to the ``cold streams''
seen in simulations (\citealt{keres_etal:05, birnboim_etal:07}).

%%%%%%%%%%%%%%%%%%%%%%%%%
% FIGURE
%%%%%%%%%%%%%%%%%%%%%%%%%
\begin{figure*}
\epsscale{1.0} 
\plotone{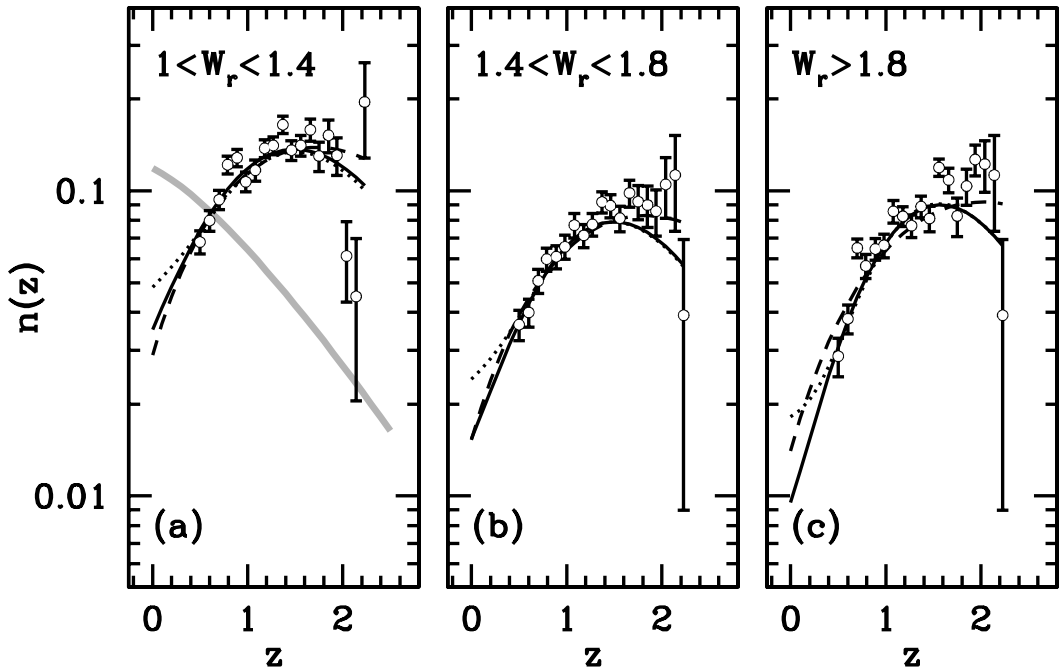}
\vspace{-5.0cm}
\caption{ \label{dndz_3panel} The redshift evolution of absorber
  counts. Points with error bars represent the measurements of
  \cite{prochter_etal:06} broken into three bins of equivalent
  width. The dashed and dotted curves represent the radius and mass
  evolution models, respectively, described in \S 2. The solid curves
  represent the combined model, discussed in \S 5. The thick gray line
  in panel (a) shows it evolution of dark matter halo counts for
  $M_h>10^{11}$ \hmsol\, multiplied by a factor of 0.01 for comparison
  to the data.}
\end{figure*}
%%%%%%%%%%%%%%%%%%%%%%%%%
% FIGURE
%%%%%%%%%%%%%%%%%%%%%%%%%

In Paper II we used an observational sample of galaxies in quasar
fields to test the assumptions and results of Paper I: namely, that
the extent of the \mg2\ gas radius is equal to $1/3$ the dark matter
halo radius, thus scaling as $M_h^{1/3}$. Brighter galaxies occupy
more massive halos, thus the implication of this assumption is that
the radial extent of \mg2\ absorption around galaxies scales with
luminosity such that

\begin{equation}
\rgas = \rgstar\left(\frac{L}{\lstar}\right)^{\alpha}.
\end{equation}

\noindent Paper I assumed that $\rgstar\approx 80$ \hkpc\ and $\alpha
\approx 0.42$, values obtained from the mean halo radius of $\lstar$
galaxies derived from halo occupation analyses of galaxy clustering at
multiple redshifts (ie, \citealt{tinker_etal:07_pvd, vdb_etal:07,
  zheng_etal:07}). These assumptions were in good agreement with the
observational sample of Paper II: $\rgstar = 91$ \hkpc\ and
$\alpha=0.35$. Once accounting for the intrinsic scaling between
galaxy luminosity and halos radius, we found the incidence of
absorption to be $>80\%$ for $L\sim \lstar$ galaxies, echoing the
result from Paper I that $\cg$ must be nearly unity at $\sim 10^{12}$
\hmsol\ to account for the frequency of absorption at $z\sim 0.6$.

In this paper, we extend the results of Papers I and II to constrain
the redshift evolution of halo occupation of the cool gas probed by
\mg2\ systems. We present two complementary models of gas-halo
evolution, demonstrating that measurements of the clustering of
absorbers at multiple epochs can provide constraints on the relative
contribution of each evolutionary scenario. We make predictions for
the evolving connection between absorbers and dark matter halos from a
single model constrained to match the same critical mass scale between
cold-mode and hot-mode halos and found in Paper I and predicted by the
hydrodynamic simulations listed above.

Throughout this paper we assume cosmological parameters in agreement
with the results of \cite{spergel_etal:07}: $\om=0.25$,
$\Omega_\Lambda=0.75$, $n=0.96$, $h=0.7$, and $\s8=0.8$. All units
assume $H_0=100\,h$ \kmsmpc. All distances are in comoving units unless
otherwise stated.

%%%%%%%%%%%%%%%%%%%%%%%%%%%%%%%%%%%
% Table 1
%%%%%%%%%%%%%%%%%%%%%%%%%%%%%%%%%%%

\begin{deluxetable*}{lccccccccccccc}
%\begin{deluxetable}{lccccccccccccccc}
\tablecolumns{14} \tablewidth{40pc} \tablecaption{Parameters of the
  Best-Fit Models} \tablehead{\colhead{Model} & \colhead{$A_W$} & \colhead{$\alpha_A$} & 
  \colhead{$\rs0$} & \colhead{$\ashock$} & \colhead{$\log \mcrit$} & \colhead{$\log \kappa_1$} &
  \colhead{$\log \kappa_2$} & \colhead{$\log \kappa_3$} & \colhead{$\log \kappa_4$} &
  \colhead{$\fcold$} & \colhead{$\beta_M$} & \colhead{$\beta_R$} & 
  \colhead{$\chi^2$} }

\startdata

Mass Evol. & 23.4 & 0.208 & 1.57 & 1.19 & 11.1 &  -1.60 & -0.445 & -0.053 & -0.114 & 0.108 & 3.51 & --- & 117.7 \\
Radius Evol. & 10.2 & 0.047 & 0.566 & 0.638 & 11.9 & -2.48 & -1.02 & -0.001 & -0.001 & 0.057 & --- & 1.47 & 122.7 \\
Combined & 14.5 & 0.197 & 1.00 & 1.00 & 11.5 & -2.00 & -0.520 & -0.130 & -0.5477 & 0.114 & 2.52 & 0.41 &122.4 \\

\enddata 

%\end{deluxetable}
\end{deluxetable*}

\section{Halo vs. Absorber Evolution}

\cite{prochter_etal:06} measured the redshift evolution of absorber
counts from Data Release 3 of the Sloan Digital Sky Survey. Figure
\ref{dndz_3panel} shows the \cite{prochter_etal:06} results broken
into three bins in equivalent width. They found that the number of
absorbers per line of sight per unit redshift, $n(z)$, increases by
nearly a factor of 3 from $0.5<z<1.5$ for $\wr>1$ systems, translating
to roughly a constant number of absorbers per unit comoving
distance. This is the opposite to the growth of dark matter structure;
as redshift increases, the amplitude of dark matter fluctuations
decreases, reducing the number of dark matter halos. The thick gray
curve in Figure \ref{dndz_3panel}a indicates $n(z)$ for dark matter
halos with $M\ge 10^{11}$ \hmsol (roughly the mass scale for $L\sim
0.1\lstar$ galaxies), weighted by the cross section of each halo. The
halo number counts decrease by more than a factor of 5 over the same
redshift range as the observations. The apparent antithetical
evolution of halos and absorbers indicates that the gaseous content of
dark matter halos must be evolving with cosmic time.

A change in the \mg2\ gaseous cross section physically corresponds to
a change in the amount and/or distribution of cool, metal-enriched gas
in the dark halo population. This could result from a change in gas
accretion onto the halo or in outflows from the galaxy within the
halo, as well as a redistribution of the metal-enriched cold gas
within the halo. In purely geometric terms, these changes could result
in an expansion or contraction of the extent of the gaseous halo per
object, or it could change the total number of halos contributing to
the overall cross section. Because the focus of our model is to
accurately constrain the halo distribution producing the absorption,
we parameterize the geometric consequences of an evolving baryonic
content without making assumptions about the mechanisms. Thus we
consider two modes of gas-halo evolution that can counterbalance the
growth of cosmic structure. First, the cross section per halo may be
changing with redshift. In this evolutionary mode, the halo mass
regime in which \mg2\ absorption originates is constant, but the cross
section per halo grows at a rate commensurate with the decreased halo
abundance. We will refer to this scenario as ``radius evolution''. In
the second evolutionary mode, the extent of the gas halo (with respect
to its dark matter halo) is constant with redshift, but the halo mass
scale evolves. In Paper I we found that $\cg$ for halos of $M\lesssim
10^{11}$ \hmsol\ falls off steeply, thus these halos contribute little
to the overall cross section at $z=0.6$. In the cold mode/hot mode
paradigm, all halo gas at this halo mass scale should be cold, thus
the decreased cross section is more likely to be explained by a change
in the distribution of cold gas or metal content than a change in the
temperature structure of the gas. We discuss a possible physical
interpretation of the behavior at low mass in \S 6.2. If this lower
mass limit decreases with increasing redshift, it would compensate for
the lower overall amplitude of the mass function at high $z$. We will
refer to this scenario as ``mass evolution''. \footnote{There is also
  the possibility that absorption efficiency, $A_W$, changes with
  redshift at fixed halo mass. This is effectively similar to radius
  evolution; increasing the strength of absorption expands the radius
  at which a given $\wr$ occurs, even though the extent of the gas
  radius is fixed. However, without any change in the gas radius
  relative to $R_{\rm halo}$, this form of evolution cannot match the
  observations; the increase in the cross section at high $\wr$ comes
  at the expense of cross section at lower $\wr$. The total number of
  absorbers integrated over all equivalent width cannot
  change. Because radius evolution entails an increase in the cross
  section per halo at fixed $\wr$, we concentrate on the first two
  scenarios.}

Absorber number counts, either in the form of $n(z)$ or the frequency
function $\fr$, cannot distinguish between these two
scenarios. However, measurements of the large-scale bias factor $b_W$
can break the degeneracy and discriminate between radius and mass
evolution. These two evolutionary processes are complementary but not
mutually exclusive; the truth may lie somewhere in between. In a third
model, we constrain the relative contributions of mass and radius
evolution with a model constrained to match the expected transition
mass scale between cold mode and hot mode accretion found in Paper
I. We will discuss this ``combined evolution'' model in detail in \S
5.

One point to stress is that dark matter halos have comoving radii that
are {\it constant} at fixed halo mass. The transition regions between
virialization, infall, and Hubble flow are determined by the
background matter density, which is constant in comoving units as
well. In physical units, dark matter halos ``shrink'' as redshift
increases, but their density relative to the background is
constant. This will become important when discussing $\cg(M,z)$ for
low-mass halos.

%%%%%%%%%%%%%%%%%%%%%%%%%
% FIGURE
%%%%%%%%%%%%%%%%%%%%%%%%%
\begin{figure*}
\epsscale{1.1} 
\plotone{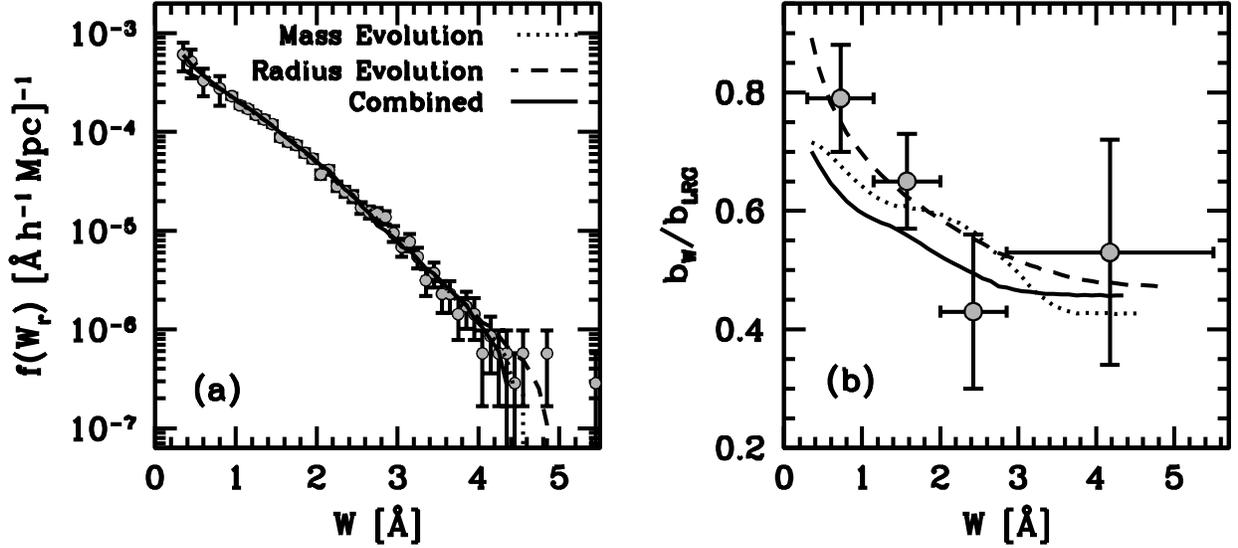}
\vspace{-9cm}
\caption{ \label{stats} Panel (a): Comparison between $\fr$ of the
  models fits and the data. Panel (b): Comparison between the
  $b_W/b_{LRG}$ at $z=0.6$ from the models and the
  \cite{bouche_etal:06} data. The models cannot be distinguished by
  frequency data and by bias data at only one redshift.  }
\end{figure*}
%%%%%%%%%%%%%%%%%%%%%%%%%
% FIGURE
%%%%%%%%%%%%%%%%%%%%%%%%%

\section{Methods}

\subsection{The Halo Occupation Model for Cold Gas}

To construct the gaseous halo, we follow the outline in Paper I with
minor modifications that we note below. First we choose the
density profile of the cool, clumpy gas that forms part of the gaseous
halo responsible for \mg2\ absorption. We assume this profile to be a
cored isothermal model of the form

\begin{equation}
\label{e.isothermal}
\rho_{\rm gas}(r) \propto \left(r^2 + a_h^2\right)^{-1}.
\end{equation}

\noindent The absorption equivalent width is proportional to the
projected density at a given impact parameter, $\pb$. This is
consistent with observations that demonstrate that \mg2\ absorbers
seen in low-resolution spectra are broken into discrete
components---cold clumps or clouds---when observed in high
resolution. The total $\wr$ is linearly proportional to the number of
components. This leads to a relationship between impact parameter and
$\wr$ of

\begin{equation}
\label{e.w_rho_propto}
\wr(\pb) \propto \frac{1}{\sqrt{\pb^2/a_h^2 + 1}}\arctan\sqrt{\frac{R_g^2 - \pb^2}{\pb^2 + a_h^2}},
\end{equation}

\noindent where $\rgas$ is the extent of the cool-gas halo. A
relationship of this form is in good agreement with the results of
Paper II. The results of Paper II also indicate that the extent of the
\mg2\ gaseous halo is $\rg\approx 0.4R_{200}$ at $z=0.6$ and
$a_h/\rg\approx 0.2$. We define the dark matter halo as having a mean
interior density 200 times the cosmic mean density, thus

\begin{equation}
\label{e.r200}
R_{\Delta} = \left(\frac{3M_h}{4\pi \rhobar\Delta}\right)^{1/3},
\end{equation}

\noindent where $\rhobar=\rhocrit\om$ and $\Delta=200$. Recall that
equation (\ref{e.r200}) implies that the {\it comoving} halo radius is
the {\it same} at all redshifts for halos of the same mass.

We normalize the $\wr(\pb,M_h)$ relation (Eq. [\ref{e.w_rho_propto}]) by
the equivalent width obtained by a sightline that passes through the
center of the halo,

\begin{equation}
\label{e.w0}
\wr(0|M_h) = A_W(M_h)\,\mtwelve^{1/3},
\end{equation}

\noindent where $\mtwelve$ is the halo mass in units of $10^{12}$
\hmsol. The function $A_W(M_h)$ encapsulates physical properties of
the gas halo, including the fraction of \mg2\ gas and the properties
of the cold clouds or clumps within the halo. If the \mg2\ gas
fraction were fixed, more massive halos would contain more \mg2\ gas
and have a larger path length at $\pb=0$, thus the natural scaling of
$\wr$ is to increase as $M_h^{1/3}$. The function $A_W(M_h)$
represents deviations from this natural scaling. Some dispersion about
$\wr(0|M_h)$ is indicated from the results of Paper II, but this
affects $\wr(\rhob)$ at small impact parameters where the cross
section is too small to make a significant impact on the
calculations. The dependence of $\wr$ on $\pb$ and $M_h$ is then

\begin{equation}
\label{e.w_rho}
\wr(\pb,M_h) = \frac{1}{\sqrt{\pb^2/a_h^2 + 1}}\arctan\sqrt{\frac{R_g^2 - \pb^2}{\pb^2 + a_h^2}}
\left[\frac{\wr(0|M_h)}{\arctan(\rgas/a_h)}\right],
\end{equation}

\noindent where the factor of $1/\arctan(\rgas/a_h)$ is to normalize
the expression, due to the fact that equation (\ref{e.w_rho_propto})
does not asymptote to unity at $\pb=0$.

Above a certain mass scale, shock heating eliminates the majority of
cold halo gas, attenuating the absorption efficiency of the halo. We
introduce a free parameter $\fcold$ to set the fraction of remaining
cold gas once the halo becomes fully shock-heated.  The resulting
$W(\pb|M_h)$ in the hot mode is simply equation (\ref{e.w_rho})
multiplied by $\fcold$. In between the cold and hot modes, gas heating
occurs as an inside-out process, resulting in a shock-heated core of
gas with core radius that increases with halo mass (see, eg,
\citealt{dekel_birnboim:06}). We parameterize the dependence of the
shock radius with mass as

\begin{equation}
\label{e.rshock}
\frac{\rshock}{\rg} = \rs0 + \ashock\log \mtwelve,
\end{equation}

\noindent and restrict $\rshock/\rg$ to be no smaller then $0$ and no
larger than $1$. Inside $\rshock$, the absorption efficiency is
reduced by $1-\fcold$. The form of equation (\ref{e.rshock}) is
motivated by the simulation results of \cite{keres_etal:05}, in which
the fraction of cold accreted gas decreases linearly with $\log M_h$
through the transition region. We will refer to the ``transition
mass'' $\mcrit$ as the mass at which $\rshock/\rg =0.5$.

It is not necessary for a sightline that passes through a halo to
produce any absorption; the clumpy nature of the cold gas may create a
covering fraction of less than unity. Alternately, only a fraction of
halos may contain extended \mg2\ gas halos. We define the product of
the covering fraction and incidence as $\cg(M_h)$, the probability
that a halo produces absorption if intersected along the line of
sight. We will define the functional form of $\cg(M_h)$ in the
following section. See Paper I for full details on the calculation of
$P(\wr,M_h)$.

\subsection{Parameterizing the dependence of $P(\wr)$ on halo mass
  and redshift}

In Paper I we set $A_W$ to be a constant with halo mass. This
assumption was influenced more by statistics than theory; although
there is no reason to expect that absorption efficiency should remain
constant with halo mass, the data used were not sufficient to
constrain a mass-dependent $A_W$. The addition of $n(z)$ ameliorates
this problem. We add an additional degree of freedom by allowing the
absorption efficiency to vary in the cold mode:

\begin{equation}
\label{e.a_w}
A_W(M_h) = \left\{ \begin{array}{ll}
    A_0\mtwelve^{\alpha_A} & {\rm if\ \ } M_h<\mcrit \\
    A_0(\mcrit/10^{12})^{\alpha_A} & {\rm if\ \ } M_h\ge\mcrit.\\
    \end{array}
    \right.
\end{equation}

\noindent We will discuss the effect of this addition in more detail
in \S 5.2. As in Paper I, we model the mass dependence of $\cg$
non-parametrically; we set the values of $\cg$ at four masses and
spline interpolate between these masses: $\log M_i = 10.0$, 11.33,
12.66, and 14.0, and refer to the values of $\cg$ at these masses as
$\kappa_1$-$\kappa_4$.

To parameterize the mass evolution mode we adjust $\cg$ as a function
of redshift. In Paper I we found that $\cg\approx 1$ at $M>10^{12}$
\hmsol, and it rapidly falls at lower masses. Thus, instead of
altering the values of $\kappa_1$-$\kappa_4$, we assume that the {\it
  shape} of $\cg(M_h)$ is fixed at all redshifts, but the {\it mass
  scale} evolves with time. In other words,

\begin{equation}
\label{e.m_evol}
M_i(z) = M_{i,0}\exp[-\beta_M(z-z_0)],
\end{equation}

\noindent where $i=1$-4, and $M_{i,0}$ is the value at $z_0=0.6$
listed in the paragraph above. We choose an exponential form for
equation (\ref{e.m_evol}) because it describes the growth rate of
individual halos well (\citealt{wechsler_etal:02,
  vdb:02}). Cluster-sized halos have high growth rates, while halos
less massive than the Milky Way have little evolution between
redshifts 1 and 0. With equation (\ref{e.m_evol}), a comparison can be
made between the growth rate required to match $n(z)$ and that of
halos of multiple mass scales.

In pure radius evolution, the values of $M_i$ are held constant at all
$z$, but the gas radius evolves as

\begin{equation}
\label{e.r_evol}
\rg(M_h,z) = R_{200}(M_h)\times 0.4\,\left(\frac{1+z}{1+z_0}\right)^{\beta_R}.
\end{equation}

\noindent Equation (\ref{e.r_evol}) implies that the gas radius can
evolve with respect to the virial radius of the dark matter. A value
of $\beta_R=1$ is equal to the gas radius being fixed in physical
units, while a value of $\beta_R=0$ means the gas halo is fixed in
comoving units, i.e., a fixed fraction of the dark matter
halo radius.

In both models we have ten free parameters: in both models there are
four that govern the covering fraction ($M_1-M_4$), two that control
the cold-hot transition ($\rs0$ and $\ashock$ from equation
\ref{e.rshock}), one for the cold fraction within the hot mode
($\fcold$), and two that control the absorption efficiency parameter
$A_W(M)$ ($A_0$ and $\alpha_A$ from equation \ref{e.a_w}). In each
model there is one parameter to govern redshift evolution ($\beta_M$
and $\beta_R$ from equations \ref{e.m_evol} and \ref{e.r_evol},
respectively). This is two more parameters than in Paper I, but we
have added 57 data points for $n(z)$ in three $\wr$ ranges, for a
total of 104 data points.

%%%%%%%%%%%%%%%%%%%%%%%%%
% FIGURE
%%%%%%%%%%%%%%%%%%%%%%%%%
\begin{figure}
\epsscale{2.3} 
\plotone{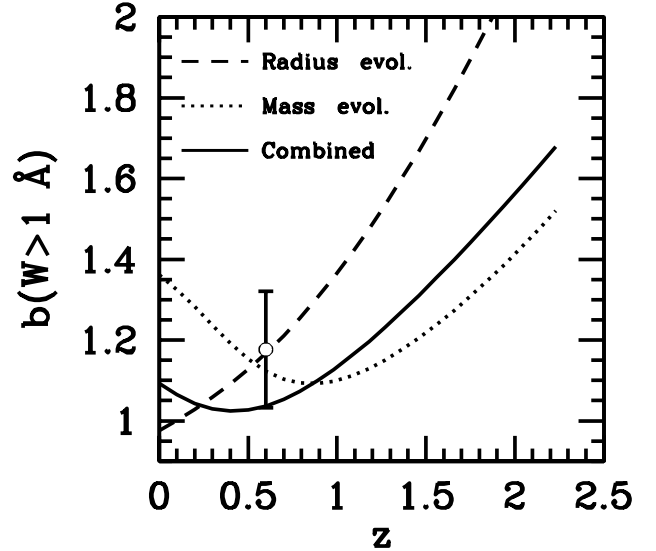}
\vspace{-9cm}
\caption{ \label{bias_redshift} Predictions for the bias of $\wr>1$
  \AA\ absorbers for each of the three evolutionary models. The data
  point at $z=0.6$ is the overall bias of absorbers from
  \cite{bouche_etal:06}. The bias curves represent the absolute bias
  with respect to the underlying dark matter distribution. }
\end{figure}
%%%%%%%%%%%%%%%%%%%%%%%%%
% FIGURE
%%%%%%%%%%%%%%%%%%%%%%%%%

%%%%%%%%%%%%%%%%%%%%%%%%%
% FIGURE
%%%%%%%%%%%%%%%%%%%%%%%%%
\begin{figure}
\epsscale{2.3} 
\plotone{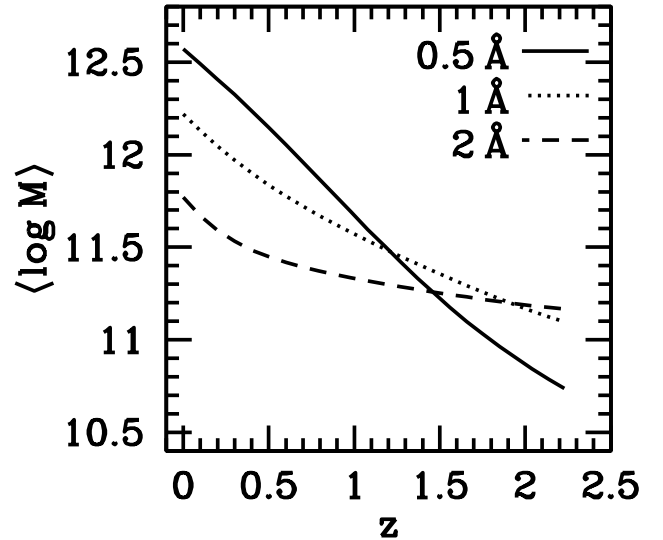}
\vspace{-9cm}
\caption{ \label{mass_evolution} The mean logarithmic halo mass for
  three different values of $\wr$. $\logmbar$ evolves strongly with
  redshift for weak absorbers but is nearly flat for strong
  absorbers. At $z<1$, this produces the $b_W$-$\wr$
  anti-correlation. At $z\sim 1.3$, $\wr$ is roughly independent of
  redshift. At higher $z$, $\wr$ and $b_W$ are positively correlated.}
\end{figure}
%%%%%%%%%%%%%%%%%%%%%%%%%
% FIGURE
%%%%%%%%%%%%%%%%%%%%%%%%%

\subsection{Calculating Observables with the Model}

Once $P(\wr|M_h,z)$ is known, the redshift evolution of absorber
counts is obtained by integrating this probability distribution
function (PDF) over the cross section-weighted halo mass function,
i.e.,

\begin{eqnarray}
\label{e.model_dNdz}
n(z) & = & \frac{c}{H_0}\left[\om(1+z)^3+\Omega_\Lambda\right]^{-1/2}\int_{W_{\rm min}}^{W_{\rm max}} d\wr \times\\ \nonumber
& & \int dM_h \left(\frac{dn}{dM_h}\right)_z \sigma_g(M_h,z) P(\wr|M_h,z),
\end{eqnarray}

\noindent where $dn/dM_h$ is the halo mass function, for which we use
the numerical results of \cite{tinker_etal:08}. The subscript $z$ on
$dn/dM_h$ indicates that the mass function is calculated at that
specific redshift. The prefactor on the right hand side converts the
linear density of absorbers from number per unit comoving distance to
number per unit redshift. We use the data from \cite{prochter_etal:06}
for $n(z)$ in three bins in equivalent width: $\wr = [1.0, 1.4]$ \AA,
$\wr = [1.4, 1.8]$ \AA, and $\wr>1.8$ \AA. To estimate the errors on
the data we add the Poisson fluctuations in quadrature with a 5\%
`systematic' error. \cite{prochter_etal:06} do not estimate the cosmic
variance of their measurements, and it can be seen in Figure
\ref{dndz_3panel} that the point-to-point variations in $n(z)$ can be
significantly larger than the error bars we have estimated, suggesting
that the true errors are underestimated.

The frequency function of absorbers is obtained by integrating over
both mass and redshift at a given $\wr$:

\begin{equation}
\label{e.model_dndW}
\fr = \frac{1}{g_{\rm tot}}\int dz\,g(z) \int dM_h \left(\frac{dn}{dM_h}\right)_z \sigma_g(M_h,z) P(\wr|M_h,z),
\end{equation}

\noindent where $\sigma_g=\pi \rgas^2$,

\begin{equation}
g_{\rm tot} = \int dz\,g(z),
\end{equation}

\noindent and $g(z)$ is the survey completeness function at each
redshift (from \citealt{prochter_etal:06}). As in Paper I, we
construct the $\fr$ data from a combination of \cite{prochter_etal:06}
for $\wr>1$ \AA, and \cite{steidel_sargent:92} for $0.3$\AA$<\wr <
1.0$\AA, where we have adjusted the normalization of the
\cite{steidel_sargent:92} data to the level of the
\cite{prochter_etal:06} results at $\wr=1$ \AA\footnote{Due to an
  error in the normalization of the \cite{prochter_etal:06} data, we
  have increased the amplitude of $\fr$ and $n(z)$ by 30\%
  (J.~X.~Prochaska, private communication). This adjustment brings
  the \cite{prochter_etal:06} data into agreement with the $n(z)$
  fitting function given in \cite{prochter_etal:06b}, which is based
  on more recent results.}. As with $n(z)$, we estimate the errors
through a quadrature addition of Poisson errors and a 5\% systematic
error bar.

The bias of absorbers is calculated in manner similar to $\fr$, but
now each halo is weighted by its large-scale bias $b(M_h,z)$ (using
the bias formula of \citealt{tinker_etal:05}).

\begin{equation}
\label{e.model_bias}
b_W(z) = \frac{1}{f(\wr,z)} \int dM \left(\frac{dn}{dM}\right)_{z} \sigma_g(M,z)\, b_h(M,z)\, P(\wr|M,z).
\end{equation}

\noindent When comparing to the data we set
$z=z_0=0.6$. \cite{bouche_etal:06} measure the bias of \mg2\ absorbers
by cross-correlating absorbers with luminous red galaxies (LRGs) from
the Sloan Digital Sky Survey photometric catalog. To compare the
models to the data, we calculate the absolute bias of absorbers from
equation (\ref{e.model_bias}) and divide the calculation by the bias
of LRGS at $z=0.6$, $b_{LRG}$ = 1.81 (\citealt{white_etal:07}).  The
bias data are in four bins of $\wr$: [0.3,1.15] \AA, [1.15,2.0] \AA,
[2.0,2.85] \AA, and [2.85,4.0] \AA. When comparing models to bias
data, we take the number-weighted average bias over the given range in
$\wr$. See Paper I for further details.

When calculating observables, we truncate the integrals over mass at
$M_h=10^9$ \hmsol. We find that extending this limit to lower masses
produces negligible differences in the model predictions, even at high
redshift. We find the best-fit models by minimizing the total $\chi^2$
from the sum of the $\chi^2$ values from $\fr$, $n(z)$, and $b_W$. We
use the Monte Carlo Markov Chain (MCMC; see, e.g.,
\citealt{dunkley_etal:05}) method to minimize $\chi^2$ and quantify
the constraints on individual parameters. Results of the fitting are
listed in Table 1.

%%%%%%%%%%%%%%%%%%%%%%%%%
% FIGURE
%%%%%%%%%%%%%%%%%%%%%%%%%
\begin{figure}
\epsscale{1.9} 
\vspace{-1.5cm}
\plotone{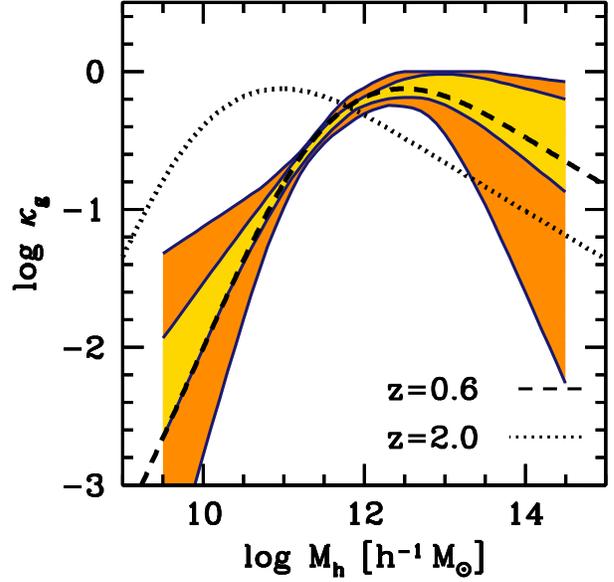}
\vspace{-5.cm}
\caption{ \label{kappa_g} Constraints on the covering factor within
  $\rg$ from the combined model. Inner and outer contours represent
  models with $\Delta\chi^2<1$ and 4 with respect to the best fit
  model, respectively. The dashed curve is the best-fit model at
  $z=0.6$, while the dotted curve represents $\cg(M)$ at $z=2.0$. At
  $z=2$, the decrease in $\cg$ above $10^{12}$ \hmsol, has little
  effect on the \mg2\ statistics because the abundance of hot-mode
  halos is significantly attenuated at this and higher redshifts.}
\end{figure}
%%%%%%%%%%%%%%%%%%%%%%%%%
% FIGURE
%%%%%%%%%%%%%%%%%%%%%%%%%

\section{Results}

Figure \ref{dndz_3panel} presents the best-fit models for both the
mass evolution model and the radius evolution model (we will discuss
the combined model subsequently). Both these models produce
statistically good fits, with $\chi^2/\nu=1.30$ for radius evolution
and $\chi^2/\nu=1.25$ for mass evolution, with the caveat that the
error bars are only estimates, and no covariance between data points
is accounted for. Figure \ref{stats} compares the model fits to the
$\fr$ and bias data. The results for $\fr$ from each model are nearly
indistinguishable, producing the observed exponential distribution of
absorbers. In our model, the exponential distribution of absorber
equivalent widths is produced primarily from the cutoff in absorber
efficiency in the hot mode. At $\wr<1$ \AA, there is evidence that
$\fr$ becomes a power law (\citealt{steidel_sargent:92,
  nestor_etal:05}). Above $\wr=2$ \AA, most all absorbers are produced
in halos in the transition stage, $M_h\approx 10^{11.5}$
\hmsol. Without a high-mass cutoff, $\fr$ is a power-law that
overproduces the abundance of high-$\wr$ systems. Below 1 \AA, the
abundance of absorbers is not attenuated by the hot mode, thus
producing a steeper $\fr$. This is not a unique solution to the
problem; it is straightforward to construct a mass-dependent covering
factor function that maps the $M_h^{-2}$ halo mass function to the
$\exp(-\wr)$. However, without the high-mass cutoff one cannot fit
both the $\fr$ and clustering data at the same time. Both evolution
scenarios produce excellent fits to $b_W$ at $z=0.6$, yielding a
strong anti-correlation with $\wr$, as shown in Figure
\ref{stats}. The amplitude of $b_W$ at $z_0$ is somewhat higher in the
radius evolution model due to the lower value of $\rs0$, which
controls the transition mass $\mcrit$. The higher mass scale for
absorbers in the radius evolution model increases the bias at all
$\wr$.

As expected, both models produce nearly identical fits to $n(z)$; the
number counts alone cannot distinguish between these models. To match
$n(z)$ in the radius evolution model, the gas radius must increase
relative to $R_{200}$ as $(1+z)^{1.47\pm
  0.05}$. \cite{prochter_etal:06} find that $n(z)$ increases roughly
as $(1+z)^{1.4}$. Because the gas cross section scales as $\rg^2$, the
redshift evolution of the cross section (at fixed halo mass) is much
larger than the increase in the number counts themselves. However,
this is counterbalanced by the reduced frequency of dark matter halos
at higher redshifts, producing an increase in absorber counts
consistent with the data. However, in order to match the number
counts, the gas radius must {\it expand in physical units} to keep the
total gas cross section constant because the number of halos is
decreasing with lookback time. At this level of evolution, the gas
radius is {\it twice} the virial radius of the dark matter halo at
$z=2$, with nearly unit covering factor and incidence.

For the mass evolution model, the growth rate of the $\cg$ mass scale
is $\beta_M=3.5\pm 0.2$. This evolution rate is much larger than that
found in dark matter halos; from \cite{wechsler_etal:02}, halos with
$z=0$ masses near $10^{12}$ \hmsol\ have growth rates of $\sim 0.6$,
while $z=0$ clusters have growth rates of $\sim 1.3$. This best-fit
value of $\beta_M$ is nearly the same as the evolution rate of the
non-linear mass scale, which approximately sets the average mass scale
of structures that are collapsing at a given epoch. The non-linear
mass scale roughly indicates the halo mass above which the halo mass
function experiences a Gaussian cutoff. The similarity between the
best-fit $\beta_M$ and the non-linear mass scale is not surprising,
given that the model is preserving the number of absorbers in a
cosmology in which the exponential cutoff in the mass function
evolves. Adding lower mass halos to the total gas cross section at the
same rate counterbalances the effect of the decreasing amplitude of
fluctuations at higher $z$.

One avenue for discriminating between these two models lies in
their constraints on the transition between cold and hot modes. In
order to fit the data, the two evolutionary models required the
hot/cold transition scale be substantially different. For radius
evolution, $\mcrit = 10^{11.9}$ \hmsol, while mass evolution produces
$\mcrit = 10^{11.1}$ \hmsol. Hydrodynamic simulations yield results in
between these two values. 

An empirical approach for discriminating between these models is the
evolution of their clustering. At different redshifts the two
evolutionary scenarios place the cold gas in different halos. Thus,
they make distinct predictions for the evolution of absorber bias. In
Figure \ref{bias_redshift}, we show $b_W$ for absorbers with $\wr>1$
\AA\ as a function of redshift for both models. The data point is
taken from \cite{bouche_etal:06}, and it has been multiplied by
$b_{LRG}$ to recover the absolute bias of the absorbers. In the radius
evolution model $b_W(z)$ increases rapidly. The mean mass scale of
absorbers in this model is nearly constant with redshift, but halos at
fixed mass correspond to increasingly rare objects at higher redshift,
resulting in increased bias. In the mass evolution model, the mass
scale decreases such that the $b_W$ at $z=1.5$ is the same as it is at
$z=0.6$. The large difference between the predictions of these two
models demonstrates the potential of discriminating between these two
types of evolution based solely on their clustering, either at high
redshift or at $z\sim 0$. Bias values in between these two extremes
would represent a combination of the two evolutionary modes. The bias
curves also diverge at low redshifts. The difference between the
models at $z=0.1$ is not much larger than the $z=0.6$ error bar, so
bias measurements at lower redshift require more precise $z=0.6$ data
to constrain models. However, ongoing spectroscopic surveys should
significantly enhance current $z=0.6$ measurements\footnote{Baryon
  Oscillation Spectroscopic Survey, http://cosmology.lbl.gov/BOSS/}.

\section{Predictions of a Combined Model}

Numerical results and analytical calculations place $\mcrit$ at $\sim
10^{11.5}$.\footnote{ We note that the canonical value of $10^{12}$
  \hmsol\ obtained by \cite{dekel_birnboim:06} specifies the mass at
  which the entire halo has been shock-heated, rather than the mass at
  which half the halo is within the shock radius, as we have defined
  it.} These results are in good agreement with the constraint on
$\mcrit$ at $z=0.6$ from Paper I. Thus, in our ``combined'' model
mentioned in \S 2, we fix the hot/cold transition to be $\mcrit =
10^{11.5}$ \hmsol, with a transition width of 1 decade in mass. With
these parameters fixed, we are able to leave both evolutionary modes
as free parameters and constrain their relative contribution to gas
halo evolution. With the transition parameters fixed to those expected
from simulations, both evolutionary models contribute, but mass
evolution dominates dominates the overall evolution of gaseous halos
($\beta_M=2.52\pm 0.24$, $\beta_R=0.41\pm 0.11$). The full parameter
set is given in Table 1.

It is important to note that there are many ways to parameterize this
problem, and the data do not currently exist that afford us the
freedom of making predictions without the use of priors on some of our
parameters. Our choices are driven by the theoretical results
discussed above. But as we demonstrated in \S 4, different choices for
these parameters result in a different halo occupation of absorbers
and result in different predictions for the clustering of absorbers
and their connection to galaxies of different luminosities. These are
predictions that further observational results will be able to test.

\subsection{The Bias of Absorbers}

The evolution of absorber bias is shown with the solid curve in Figure
\ref{bias_redshift}. Because mass evolution dominates this model, the
slope of bias with redshift is slightly steeper than from mass
evolution alone. Figure \ref{mass_evolution} shows the mean
logarithmic mass of absorbers as a function of redshift. For weak
absorbers, $\wr=1$ \AA, $\logmbar$ is a strongly decreasing function
of $z$. Strong absorbers, however, predominantly arise at $M_h\sim
\mcrit$ regardless of redshift; in the cold mode, $\wr \sim
A_wM_h^{1/3}$, thus the strongest absorbers arise in the most massive
halos before the transition to shock heating. For $\wr\sim 2$ \AA\
absorbers, $\logmbar\approx 11.3$ and it is nearly independent of
redshift at $z>0.5$.

At $z=0.6$, $\wr$\ and $\logmbar$ are anti-correlated, yielding the
$b_W$-$\wr$ anti-correlation measured. From Figure
\ref{mass_evolution} it is apparent that the anti-correlation goes
away at $z=1.3$, and at $z>2$ $\wr$ and $\logmbar$ (and also $b_W$)
are positively correlated. At $z>2$, the abundance of halos above
$\mcrit$ is negligible, thus they no longer contribution to the
clustering statistics of absorbers. The predictions for pure mass
evolution are similar to those in Figure \ref{mass_evolution}, but for
pure radius evolution the predictions are markedly different. In
radius evolution, the shape of the $b_W$-$\wr$ anti-correlation is
preserved at all redshifts. Thus, another method for discriminating
between the different evolutionary modes is bias as a function of
$\wr$ at high redshift.

\subsection{Incidence and Covering Factor}

Figure \ref{kappa_g} shows the constraints on the covering
fraction/incidence rate of \mg2\ absorbers as a function of host-halo
mass, $\cg$. At $z=0.6$, halos above $\sim 10^{12}$ \hmsol\ are
consistent with covering factor near unity. Below this mass, the
covering factor decreases roughly linearly with halo mass. The softer
cutoff at low mass is somewhat different from the constraints on
$\cg(M_h)$ in Paper I but are consistent with the observational
results of Paper II.  In Paper I, $\cg$ for $M<10^{11.5}$ decreased
rapidly, becoming essentially zero at $\sim 10^{11}$ \hmsol. This
result was driven in part by the parameterization used in Paper I, in
which $A_W$ was set to be a constant for halos in the cold mode. Here
we allow $A_W$ to vary as a power law, and in the best-fit model
$A_W\sim M^{0.2}$. Thus lower mass halos have somewhat lower
absorptions efficiency and may have a non-zero covering fraction
without overproducing the frequency of absorbers at low-$\wr$.

In the hot mode, $\cg\gtrsim 0.4$ at $M_h\gtrsim 10^{13}$ \hmsol, the
constraints are poor on the covering fraction, although the 1-$\sigma$
lower limit is $\cg\approx 10\%$. The bias data drives the constraints toward
higher $\cg$ in this mass range; the models that reproduce the
clustering anti-correlation all have $\cg\gtrsim 0.4$. To match the
strong clustering of $\wr\sim 1$ \AA\ absorbers, as well as the high
overall clustering of absorbers, a significant fraction of systems
must be produced in massive halos. However, the current level of
precision on the bias data are not sufficient to place tight
constraints on $\cg$ in the hot mode. A high incidence of absorption
at this mass scale may seem difficult to reconcile with the fact that
the cold fraction has been reduced by $\sim 90\%$ in our model, but
the model used here may be overly simplistic. In multiphase models of
gaseous halos (ie, \citealt{mo_jordi:96, maller_bullock:04}),
cold gas in hot halos extends out to the virial radius, either in the
form of condensed clouds or cold streams. Thus, it is possible to
match the data with a much lower $\cg$ in the hot mode but now with
$\rgas \approx R_{200}$. The model predictions would be nearly
identical for the current data, but the LRG-\mg2\ cross-correlation
function would be markedly different at $r\lesssim 1$ \hmpc\
scales. More study of the incidence and radial extent of cold gas
around clusters or LRGs is required to constrain the cold gas around
massive objects.

%%%%%%%%%%%%%%%%%%%%%%%%%
% FIGURE
%%%%%%%%%%%%%%%%%%%%%%%%%
\begin{figure}
\epsscale{1.6}
\plotone{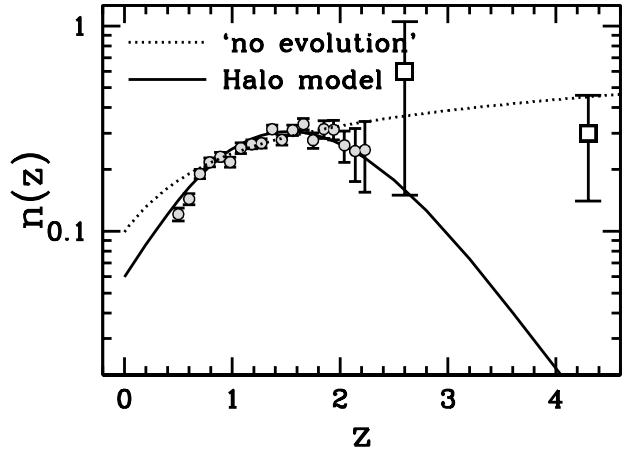}
\vspace{-5cm}
\caption{ \label{dndz} $n(z)$ for the combined evolution model,
  extrapolated to $z\gtrsim 4$. The circles are the
  \cite{prochter_etal:06} data for $\wr>1$ \AA\ ($z=2.6$) and
  $\wr>1.5$ \AA\ ($z=4.3$). The squares are the results of
  \cite{jiang_etal:07}. The dotted line is the `no evolution' model,
  in which the number of absorbers per comoving unit distance is
  constant.}
\end{figure}
%%%%%%%%%%%%%%%%%%%%%%%%%
% FIGURE
%%%%%%%%%%%%%%%%%%%%%%%%%

%%%%%%%%%%%%%%%%%%%%%%%%%
% FIGURE
%%%%%%%%%%%%%%%%%%%%%%%%%
\begin{figure*}
\epsscale{1.1}
\plotone{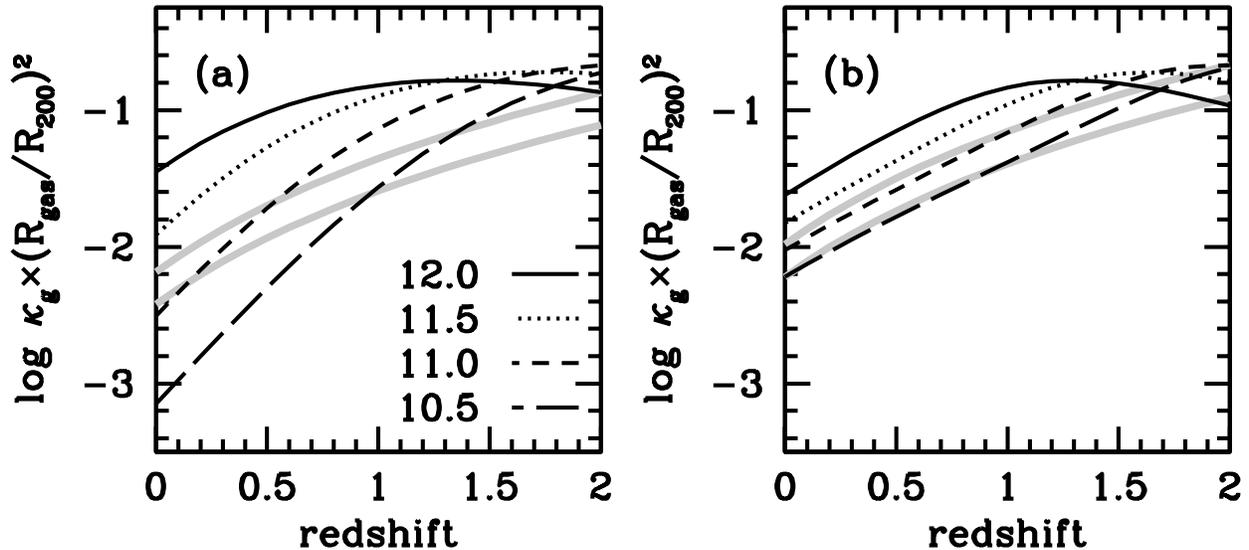}
\vspace{-9cm}
\caption{ \label{kg_evolution} Panel (a): Redshift evolution of the
  total gas cross section, $\ktot=\cg\,(\rg/R_{200})^2$, for the
  best-fit combined model. The black curves represent different halo
  masses. The key indicated $\log M$ for each line type. The two gray
  curves are the toy model results, discussed in \S 6.2, for $\log M =
  11.0$ and $10.5$ (upper and lower curves, respectively). Panel (b):
  Same as panel a, but results are shown for the $2\sigma$ upper limit
  on $\kappa_1$ shown in Figure \ref{kappa_g}. Gray curves have been
  shifted up to compare with the halo model results.}
\end{figure*}
%%%%%%%%%%%%%%%%%%%%%%%%%
% FIGURE
%%%%%%%%%%%%%%%%%%%%%%%%%

\section{Discussion}

We have presented a model that connects \mg2\ absorbers to dark matter
halos in order to reproduce their observed statistical properties;
their clustering bias at $z=0.6$, their frequency function, and their
redshift evolution. In a model in which the transition between
cold-gas halos and hot-gas halos is fixed to be $10^{11.5}$ \hmsol, we
find that the mass scale of absorbers must significantly decrease with
increasing redshift in order to match the data. Although the mass
scale of strong absorbers varies little with redshift, the overall
mass scale of absorbers is maximal at $z=0$. This upsizing is in stark
contrast to the `downsizing' trend in the star formation of galaxies,
in which the most massive systems form their stars early and the halo
mass scale at which most the star formation is occurring increases
with redshift. This model makes testable predictions of the clustering
of absorbers at $1<z<2$ and at $z\sim 0.1$.

\subsection{The evolution of absorber number counts}

The hot/cold transition is important in determining the slope of
$n(z)$ and how that slope depends on $\wr$. \cite{nestor_etal:05} find
that the slope of $n(z)$ depends strongly on equivalent width; for
$\wr\sim 0.3$ \AA, $n(z)$ is nearly constant, and the slope
monotonically increases with absorption strength. The hot/cold
transition, combined with the mass evolution scenario, offers a
succinct explanation of this effect. Strong absorbers predominantly
exist in the cold mode, thus their frequency is enhanced as the
fraction of absorbers contributed by $10^{11}$-$10^{12}$ \hmsol\ halos
increases from $z=0.5$ to 1.5. Weak absorbers are produced in {\it
  both} hot and cold halos, so the fraction of halos in the hot mode
has less relevance. The frequency of $0.3\le \wr\le 1$ \AA\ absorbers
in our combined model is nearly independent of redshift, in agreement
with the \cite{nestor_etal:05} measurements. For pure radius
evolution, the shape of $n(z)$ is independent of $\wr$, lending
further support to mass evolution as the dominant scenario.

Although the \cite{prochter_etal:06} $n(z)$ data only extend to $z\sim
2.2$, it is interesting to extrapolate the models to higher
redshifts. Figure \ref{dndz} shows the prediction for the combined
model at $z>2$.  The reduced abundance of $M\gtrsim 10^{11}$ \hmsol\
dark matter halos at high redshift eventually wins out over either
mass or radius evolution; the model peaks at $z\sim 1.5$, then turn
over and decreases monotonically at higher redshift. Lower-mass halos,
$M\lesssim 10^{9-10}$ \hmsol, still contain significant total cross
section, but in our model the absorption efficiency of these halos is
too low to produce a large cross section at $\wr\ge 1$ \AA.  The
results from the mass- and radius-evolution models are quantitatively
similar. The dotted curve in Figure \ref{dndz} is commonly referred to
as a model with `no evolution', meaning that the comoving number
density of \mg2-absorbing objects is constant and that the gas cross
section per object is constant in physical units. For this model,
$n(z) \propto
(1+z)^2[\Omega_m(1+z)^3+\Omega_\Lambda]^{-1/2}$. \cite{jiang_etal:07}
have an estimate of $n(z)$ at $z=2.6$ ($\wr>1 \AA$) and $z=4.3$
($\wr>1.5 \AA$) from a sample of six $z\sim 6$ quasars with a total
redshift pathlength of $\Delta z=0.4$. These data are preliminary; the
mean number density of \mg2\ absorbers may change when surveys of a
larger sample of QSO sightlines become available.  The current data
are consistent with the no-evolution model. But from Figure \ref{dndz}
it is clear that the no-evolution model corresponds to significant
evolution with respect to the underlying halo population, both in
terms of the typical halo mass and the size of the gas halo with
respect to the dark matter halo. When extrapolating the halo model,
the predictions are an order of magnitude below the measurements at
$z=4.3$. Removing the critical mass scale and allowing all halos to
have high absorption efficiency at this redshift (as found in the
analytic model of \citealt{dekel_birnboim:06} and the hydrodynamic
simulations of \citealt{keres_etal:08}) will not alleviate this
discrepancy due to the paucity of halos above $\mcrit$ at this epoch.

There are two major possibilities for resolving this discrepancy; the
\mg2\ cross-section around galaxy-type dark matter halos increases
significantly beyond their virial radius, or low-mass (sometimes
called `minihalos') have enhanced absorption efficiency at high
redshift. The former scenario could be the result of star formation in
typical high-redshift galaxies, where metal-enriched gas is
efficiently dispersed into the IGM to increase the cross-section of
absorbing gas. The latter scenario is the proposal of
\cite{abel_mo:98} to explain the high frequency of Lyman-limit systems
at $z>2$. Star formation in these halos is not possible after
reionization, but before that time it may be possible to form small
amounts of stars and enrich the minihalos. Further data on the
equivalent width distribution and absorber-galaxy pairs will be able
to discriminate between these two models.

\subsection{Evolution of absorbers in the cold mode}

Our combined model predicts that the overall halo mass scale of \mg2\
absorbers reduces as redshift increases. This is caused both by the
loss of massive halos as redshift increases and the the increased
gaseous cross section of low-mass halos in the cold mode. The rate of
change of the halo mass scale for low-$\wr$ systems is significantly
faster than the growth rate of low-mass halos themselves; i.e., a
$10^{11}$ \hmsol\ halo at $z=1.5$ that produces an absorber is much
less likely to produce an absorber when that halo evolves to $z=0$
even though that halo is, on average, twice as massive.

Figure \ref{kg_evolution}a shows the evolution in the mean covering
fraction within halos relative to the total halo cross section,
$\ktot=\cg\times(R_g/R_{200})^2$, for the combined model. At $10^{12}$
\hmsol, this quantity does not change much with lookback time, but at
lower masses the cross section at $z=0$ is below 1\% and the evolution
with redshift is steep.

We can try to understand the mass and redshift trends seen in Figure
\ref{kg_evolution} through a simple picture of gas accretion. The
evolution of $\cg$ occurs in the cold mode---halos less massive than
$M_h\sim 10^{11.5}$ \hmsol. At all mass scales, the accretion of
baryonic matter is expected to increase with redshift. A convenient
fitting formula is

\begin{equation}
\mgdot\approx 6.6 \mtwelve^{1.15}(1+z)^{2.25}
\end{equation}

\noindent (\citealt{dekel_etal:08}). Since there is no hot gas halo at
these masses, and all gas is accreted cold, the infalling gas will
simply fall to the center of the halo. We assume that it takes a
free-fall time, $t_{ff}$, for the gas to reach the galaxy at the
center of the halo---although the gas is accreted with non-zero
velocity, it will also have angular momentum that must be shed before
it can reach the galaxy. The important assumption is that the infall
time is related to the dynamical time of the halo. The total amount of
gas in a cold-mode halo is $M_g = t_{ff}\mgdot$. Using

\begin{eqnarray}
t_{ff} & \approx & 0.5(G\bar{\rho}_h)^{-1/2} \nonumber \\
 & \approx & 1.4\,(1+z)^{-3/2}\,{\rm Gyr},
\end{eqnarray}

\noindent the total halo gas scales as $M_g\sim
(1+z)^{0.75}M_h^{1.15}$. The free fall time does not depend on halo
mass because all halos have the same mean density of 200 times the
background, which scales as $(1+z)^3$. We are less interested in the
absolute scaling of the covering factor in this toy model than the
scaling with mass and redshift, so we can normalize this relation to
give the proper covering fraction of $10^{11}$ \hmsol\ halos at
$z=0.5$, $\ktot=0.02$. Provided that the mean properties of the
clumps---their sizes and densities---stay relatively constant, the
covering factor of the accreted gas should scale with the amount of
material and the volume throughout which they are distributed.

Not only does the gas accretion rate increase with redshift, but the
physical volume in which they are spread is decreasing---recall that
halos at fixed mass have constant comoving radii. Thus the covering
factor scales as

\begin{eqnarray}
\ktot & \sim & M_g(M_h,z)R_{200}(M_h,z)^{-2}\nonumber \\
 & \sim & M_h^{1.15}(1+z)^{0.75}M_h^{-2/3}(1+z)^{2} \nonumber \\ 
& \sim & M_h^{0.5}(1+z)^{2.75},
\end{eqnarray}

\noindent where $R_{200}$ is now specified in physical units. This
scaling has the proper behavior required to match the absorber
statistics considered here: in the cold mode, the covering factor
decreases with decreasing mass and increases with redshift. The two
gray lines in Figure \ref{kg_evolution} show this scaling relation for
$10^{11}$ and $10^{10.5}$ \hmsol\ halos, once again normalized to the
model value for $10^{11}$ \hmsol\ halos at $z=0.5$. For the best-fit
model, the mass and redshift dependence is stronger than in our toy
model.

The variables of the model that control the evolution of $\ktot$ in
the cold mode are $\kappa_1$, $\kappa_2$, which set the slope of $\cg$
as a function of halo mass, and $\beta_M$. Due to the low contribution
to the overall gas cross section from these masses, we are not able to
constrain $\kappa_1$ and $\kappa_2$ tightly. Figure
\ref{kg_evolution}b shows $\ktot(z)$ for the same halo masses but for
the $2\sigma$ upper limit on $\kappa_1$ shown in Figure
\ref{kappa_g}. In this model, $\cg$ scales as $M_h^{0.5}$ at low
masses rather than $M_h$ as in the best-fit model. In this
realization of our halo model, the scalings of $\ktot$ are in good
agreement with the expectations from the toy model. To better
constrain these parameters, tighter bias measurements are required---
or bias measurements at higher redshift---as well as more precise
measurements of $\fr$ at $\wr<1$ \AA.

The presence of an absorber in $M_h\lesssim \mcrit$ halos is indelibly
linked to the rate of gas accretion. Even if absorbers are caused by
active star formation episodes, such episodes cannot occur if there is
no fuel coming into the galaxy. If \mg2\ absorbers are related to
star-forming objects it is tempting link the decrease in the cosmic
star formation rate from $z=1$ to $z=0$ with the measured $n(z)$ for
absorbers. Certainly this is the case to some extent; the redshift
evolution of \mg2\ systems is strongest for the highest equivalent
widths. But the lack of significant evolution in $\wr<1$ \AA\
absorbers argues against such a link for all systems.  In addition,
$n(z)$ is a cross-section weighted statistic and some care is required
for quantitative comparison with the observed star formation rate
density.

\subsection{Cold gas in hot halos}

In the combined model, $\fcold$ is $\sim 10\%$. Nominally, this
parameter relates to the fraction of cold gas that remains after the
halo has been shock heated. In addition to the bias anti-correlation,
there are several additional pieces of observational evidence that
point to cold gas existing in extended hot gas halos.  First, the
incidence of \mg2\ absorbers along lines of sight near $z\sim 0.7$
galaxy clusters is significantly enhanced relative to the field
(\citealt{lopez_etal:08}). The \cite{lopez_etal:08} results favor
enhancement of stronger absorbers ($\wr\sim 2$) over weaker absorbers,
but absorption of any kind reveals the presence of cool gas within the
cluster virial radius. Second, the recent measurements of the
LRG-\mg2\ cross-correlation by \cite{lundgren_etal:09} and
\cite{gauthier_etal:09} extend the previous measurement to small
scales--- $r<1$ \hmpc. Pairs at this scale can only be created if LRGs
and absorbers exist within the {\it same halos}; ie, the ``one-halo''
term from halo models of clustering. Halo occupation analysis of the
LRG autocorrelation function constrains LRG halo masses to be $\gtrsim
10^{13}$ \hmsol\ (\citealt{padmanabhan_etal:08, blake_etal:08,
  zheng_etal:08}). The multiphase cooling model of
\cite{maller_bullock:04} predicts that cold clouds are still able to
form and survive at halos with virial velocities of $300$-$400$ \kms,
with a cold fraction of 10\% at $\sim 320$ \kms. High-resolution
hydrodynamic simulations by \cite{kaufmann_etal:08} support the
scenario in which cloud formation is efficient in hot-gas halos with
cored density profiles, consistent with the density profile of
hot-mode halos in the \cite{keres_etal:08} results, which have lower
resolution but cosmological volumes.

Detailed analysis of the cross-correlation function of LRGs and \mg2\
absorbers at both large and small scales can constrain both the
covering fraction and density profile of cold gas in hot-mode halos,
and this is the subject of ongoing study (J.-R.~Gauthier et.~al., in
preparation).

\section{Summary}

In this paper, we have used the frequency and abundance of dark matter
halos to construct a model for the frequency, abundance, and
clustering of \mg2\ absorption systems. One of the main precepts on
which our model is based is the idea of a critical mass scale, below
which absorption is high-efficiency and above which absorption is
low-efficiency but non-zero. The free parameters of this model are
constrained to match the known absorber statistics, allowing us to
constrain the relationship between absorption equivalent width and
halo mass. Our main findings are:

%\noindent 
(1) In order to match the increasing frequency of absorbers
with redshift, $n(z)$, the connection between absorbers and dark
matter halos must evolve significantly from $z=0.5$ to $z=2$. Either
gaseous halos must expand with respect to their dark matter radii, or
the halo mass range probed by absorbers must expand to account for the
data. Different evolutionary scenarios make distinct predictions for
the clustering of absorbers at both $z=0$ and $z\gtrsim 1$.

%\noindent 
(2) If $\mcrit$ remains constant at $10^{11.5}$ \hmsol\ at
all redshifts, the dominant evolutionary mechanism is in the mass of a
typical absorber. Strong absorbers are preferentially created in
$\mcrit$ halos at all redshifts, while weak absorbers ($\wr\lesssim 1$
\AA) evolve from occupying $10^{13}$ \hmsol\ halos at $z\lesssim 0.5$
to $10^{11}$ \hmsol\ halos at $z\gtrsim 1.5$.

%\noindent 
(3) In our best-fit model, the cross section of $M\lesssim
10^{11}$ increases significantly from $z=0$ to $z=2$. The scaling of
$\kappa_g(M)$ with mass and redshift is consistent with a simple
picture in which all gas is accreted cold and sinks to the center of
the halo in a freefall time. The increased covering fraction results
from higher gas accretion rates and smaller volumes of halos at high
redshift.

%\noindent 
(4) At $z=4$, our fiducial model is not able to reproduce the
preliminary measurements of the high frequency of absorbers. If these
data do not change with greater statistics, this implies that the
absorption efficiency of low-mass $M\lesssim 10^{10}$ \hmsol\ is
enhanced at these redshifts, or the cross-section around typical
galactic halos is expanded beyond their virial radii.

\vspace{1cm}

JT would like to thank Ari Maller for useful conversations. JT
acknowledges the use of the computational facilities at the Kavli
Institute for Cosmological Physics at the University of
Chicago. H.-W.C. acknowledges partial support from NASA Long Term
Space Astrophysics grant NNG06GC36G and an NSF grant AST-0607510.

%%%%%%%%%%%%%%%%%%%%%%%%%%%%%%%%%%%%%%%%%%%%%%%%%%%%%%%%%%%%%%%%%%%%%%%%
%  Bibliography
%%%%%%%%%%%%%%%%%%%%%%%%%%%%%%%%%%%%%%%%%%%%%%%%%%%%%%%%%%%%%%%%%%%%%%%%

\bibliography{../risa}

\end{document}